\documentclass{article}
\usepackage[utf8]{inputenc}
\usepackage{amsmath}
\usepackage[left=1.50cm,
right=1.50cm,
top=2.00cm,
bottom=2.00cm
]{geometry}
\usepackage{graphicx}
\usepackage{float}
\usepackage[sorting=none]{biblatex}
\usepackage{bbold}
\usepackage{caption}
\usepackage{subcaption}
\usepackage{appendix}
\usepackage{braket}
\usepackage{comment}
\usepackage{multicol}
\usepackage{multirow}
\usepackage{comment}
\usepackage{xcolor}
\usepackage{authblk}

\begin{document}

\title{Comparing coherent and incoherent models for quantum homogenization}
\author[1]{Anna Beever\thanks{annabeever1@gmail.com}}
\author[1,2]{Maria Violaris}
\author[1]{Chiara Marletto}
\author[1]{Vlatko Vedral}

\affil[1]{\small Clarendon Laboratory, University of Oxford, Parks Road, Oxford OX1 3PU, United Kingdom}
\affil[2]{Mathematical Institute, University of Oxford, Woodstock Road, Oxford OX2 6GG, United Kingdom}
\normalsize

\date{\today}

\maketitle

\begin{abstract}
    Here we investigate the role of quantum interference in the quantum homogenizer, whose convergence properties model a thermalization process. In the original quantum homogenizer protocol, a system qubit converges to the state of identical reservoir qubits through partial-swap interactions, that allow interference between reservoir qubits. We design an alternative, incoherent quantum homogenizer, where each system-reservoir interaction is moderated by a control qubit using a controlled-swap interaction. We show that our incoherent homogenizer satisfies the essential conditions for homogenization, being able to transform a qubit from any state to any other state to arbitrary accuracy, with negligible impact on the reservoir qubits' states. Our results show that the convergence properties of homogenization machines that are important for modelling thermalization are not dependent on coherence between qubits in the homogenization protocol. We then derive bounds on the resources required to re-use the homogenizers for performing state transformations. This demonstrates that both homogenizers are universal for any number of homogenizations, for an increased resource cost.
\end{abstract}

\begin{multicols}{2}
\section{Introduction}

The role of quantum effects in thermodynamics has led to many fruitful results in recent years, with several new phenomena arising from underlying quantum dynamics \cite{goold2016role}. Furthermore, a major focus for the field of quantum thermodynamics is modelling thermalization processes \cite{mori2018thermalization, kaufman2016quantum}. These models are often based on collision models with weak coupling \cite{tabanera2022quantum, ciccarello2022quantum, arisoy2019thermalization}, a very general version of which was proposed in \cite{2002paper}, using a partial-swap (\textsc{pswap}) quantum homogenizer. There it is shown that a weak \textsc{pswap} interaction between a system qubit and each identical qubit in a large reservoir will cause the system qubit to converge to the reservoir qubits' state, while leaving the reservoir qubits approximately unchanged. The homogenizer is universal in that it will transform any state to any other state, and the \textsc{pswap} is shown to be the unique operation that satisfies the homogenization conditions. Since its proposal, many investigations of thermalization have been based on the quantum homogenizer model or variations of it \cite{scarani2002thermalizing, strasberg2017quantum}.

In the \textsc{pswap} homogenizer, each interaction between the system qubit and the reservoir qubits is unitary, resulting in a web of interference between system and reservoir qubits that have interacted. This raises the question as to how far the properties of the homogenizer are affected by the coherence of the unitary \textsc{pswap}, and whether the homogenizer's properties result from non-trivial quantum phenomena. 

Here we propose a new universal quantum homogenizer, which is an incoherent variation of the \textsc{pswap} homogenizer. We introduce an additional control qubit for each reservoir qubit in the protocol, and replace the \textsc{pswap} by a controlled-swap (\textsc{cswap}) interaction, conditioned on the control qubit with a system and reservoir qubit as targets. The \textsc{cswap} gate has been previously investigated in a variety of contexts, including studies on comparing entangling power of \textsc{pswap} and controlled unitary gates \cite{balakrishnan2008entangling}, experimental implementations (e.g. \cite{ono2017implementation}), comparing quantum states, and detecting entanglement \cite{foulds2021controlled}.

Mediating the interaction via a control qubit prevents interference between the system and reservoir qubits. We place an upper bound on the difference between the system qubit convergence achieved using the \textsc{cswap} and \textsc{pswap} homogenizers for arbitrary system and reservoir states, demonstrating that the difference tends towards zero as the size of the homogenizer increases. Furthermore, we identify a number of cases where the homogenizations are identical. We reinforce our conclusions with numerical simulations. Our analysis shows that the states in the two protocols differ in their paths to converging to a state, and also have major differences in the joint entropy of the system and environment qubits, but these aspects do not affect the homogenization properties. 

In addition, we derive new results regarding the reusability of both homogenizers. By calculating lower bounds on the resources needed to homogenize a general number of system qubits, we conclude that there always exists a protocol for approximately homogenizing $n$ system qubits to within a given error $\Delta$, with $N$ reservoir qubits remaining $\Delta$ close to their initial state. This requires making $N$ larger and the coupling strength weaker than the equivalent constraints for performing only a single homogenization within some error. Our analysis of the \textsc{cswap} is more general than that for the \textsc{pswap} as the lack of coherent terms simplifies the analysis, leading to tighter bounds for that protocol. 

Our results also suggest that recently found results about a new form of irreversibility and information erasure in quantum homogenization machines (see \cite{violaris2022irreversibility, marletto2022emergence}) are not dependent on non-trivial quantum coherence in the homogenizer, making them more generally applicable than was previously shown. 

\subsection{PSWAP quantum homogenizer}
    
    The quantum homogenizer was originally proposed as a model for thermalization \cite{2002paper, scarani2002thermalizing}. It consists of a set of identical reservoir qubits, which each interact one by one with a system qubit, via a unitary \textsc{pswap} interaction:
    
    \begin{equation} \label{partial swap}
    U = \text{cos}\eta \mathbb{1} + i\text{sin}\eta \mathbb{S}.
    \end{equation}

    The \textsc{pswap} is a combination of the identity $\mathbb{1}$ and the SWAP $\mathbb{S}$, weighted by a coupling strength parameter $\eta$. The system qubit converges to the state of the reservoir qubits as the size of the reservoir $N$ is increased, meanwhile the reservoir qubits stay arbitrarily close to their original state as the coupling strength of the \textsc{pswap} is made small. Hence the quantum homogenizer approximately erases the state of the system qubit, such that all reservoir qubits and the system qubit are close to the original state of the reservoir qubits. Specifically, it implements the following transformation: 

    \begin{equation}
    U^{\dagger}_N...U^{\dagger}_1 (\rho\otimes \xi^{\otimes N})U_1 ...U_N\approx\xi^{\otimes N+1}
    \end{equation}

    where $U_k := U \otimes (\otimes_{j\neq k} \mathbb{1}_j)$ denotes the interaction between the system qubit, which begins in the state $\rho$ and the $k^{\text{th}}$ reservoir qubit, which begins in the state $\xi$. 

    There are two conditions that must be satisfied for homogenization. For any distance $\delta$, defined according to some distance measure between quantum states such as trace norm, the system qubit must become at least $\delta$ close to the initial reservoir qubit state, with all the reservoir qubits also at least $\delta$ close to their initial state. Formally, for some distance measure $D(\rho_1, \rho_2)$ and number of reservoir interactions $N$: 

    \begin{equation}
        D(\rho_N,\xi) \leq \delta
    \end{equation}
    and
    \begin{equation}
        D(\xi_j,\xi) \leq \delta ~ \forall ~ j, j \leq N
    \end{equation}
    
    for arbitrarily small $\delta$. Here $\rho_{j}$ denotes the state of the system qubit after interacting with $j$ reservoir qubits, and the $j^{\textrm{th}}$ reservoir qubit to interact with the system is denoted by $\xi_{j}$.

    It was shown in \cite{2002paper} that the quantum homogenizer based on the \textsc{pswap} satisfies these conditions, for any initial state of the system and reservoir qubits. Furthermore, the \textsc{pswap} is the only unitary operator that satisfies the conditions, meaning it uniquely determines the universal quantum homogenizer. 

\section{CSWAP quantum homogenizer}\label{incoherent}

    We will now define a universal quantum homogenization protocol based on the \textsc{cswap} instead of the \textsc{pswap}, removing the coherence between the system qubits and reservoir qubits. The \textsc{cswap} operation is a three-qubit gate, where the two-qubit swap operation is applied to the 2$^{\textrm{nd}}$ and 3$^{\textrm{rd}}$ qubits if the 1$^{\textrm{st}}$ (control) qubit is a $\ket{1}$, and they are left alone if the control qubit is a $\ket{0}$: 

    \begin{equation} \label{control swap}
    U = \frac{1}{2} ( \ket{0}\bra{0} \otimes \mathbb{1} + \ket{1}\bra{1} \otimes  \mathbb{S} ).
    \end{equation}
    
    In our protocol, the control qubit begins in the state $\ket{c}$, a weighted superposition of $\ket{0}$ and $\ket{1}$, parameterized by a coupling strength $\eta$:
    
    \begin{equation}
        \ket{c} = \cos \eta \ket{0} + \sin \eta \ket{1}
    \end{equation}
    
    Consider a system qubit and a reservoir qubit with initial states $\rho_0 = \frac{\mathbb{1} + \vec{\beta} \cdot \vec{\sigma}}{2}$ and $\xi_1 = \frac{\mathbb{1} + \vec{\alpha} \cdot \vec{\sigma}}{2}$ respectively. Here the subscript zero indicates that the system qubit has interacted with zero reservoir qubits. Table \ref{state_table1} shows the results of letting the two qubits interact with a control qubit $c$, initially in the state $\rho_c = \ket{c}\bra{c}$, via the \textsc{cswap} interaction. The table shows the final joint state of the system and reservoir qubits, and the final state of the system qubit. Table \ref{state_table2} shows the corresponding states when the two qubits instead interact via a \textsc{pswap} interaction. The key difference between the final joint states in the two cases is that there are additional terms in the final joint state and final state of the system qubit when the \textsc{pswap} is used instead of the \textsc{cswap}. These additional terms indicate coherence between the qubits, and by comparing the \textsc{pswap} and \textsc{cswap} we investigate how far they impact the convergence and reusability properties of a quantum homogenization protocol.

        \begin{table}[H]
    \caption{Initial and final states after \textsc{cswap}.}
    \centering
    \begin{tabular}{ |c|c| }
    
     \hline
      & Controlled Swap \\ 
     \hline
     Initial States & $\rho_0 = \frac{\mathbb{1} + \vec{\beta} \cdot \vec{\sigma}}{2}$, $\xi = \frac{\mathbb{1} + \vec{\alpha} \cdot \vec{\sigma}}{2}$, $\rho_c = \ket{c}\bra{c}$ \\ 
     \hline
     Final Joint State & $\rho_{\textrm{s+r}}^{\textsc{cswap}} = c^2 (\rho_0 \otimes \xi) + s^2 (\xi \otimes \rho_0)$ \\ 
     \hline
     Final System State & $\rho_1 = \frac{\mathbb{1}}{2} + \frac{s^2}{2} \vec{\alpha} \cdot \vec{\sigma} + \frac{c^2}{2} \vec{\beta} \cdot \vec{\sigma}$\\
     \hline
    \end{tabular}
    \label{state_table1}
    \end{table}
    
    \begin{table}[H]
    \caption{Initial and final states after \textsc{pswap}.}
    \centering
    \begin{tabular}{ |c|c| }
    
     \hline
      & Partial Swap \\ 
     \hline
     Initial States & $\rho_0 = \frac{\mathbb{1} + \vec{\beta} \cdot \vec{\sigma}}{2}$, $\xi = \frac{\mathbb{1} + \vec{\alpha} \cdot \vec{\sigma}}{2}$ \\ 
     \hline
     Final Joint State & $\rho_{\textrm{s+r}}^{\textsc{pswap}} = c^2 (\rho_0 \otimes \xi) + s^2 (\xi \otimes \rho_0)$ \\
      & - $\frac{cs}{8} (\vec{\beta} - \vec{\alpha}) \cdot (\sigma \otimes \mathbb{1} \wedge \mathbb{1} \otimes \sigma)$ \\
      & - $\frac{cs}{8} (\vec{\beta} \wedge \vec{\alpha}) \cdot (\sigma \otimes \mathbb{1} - \mathbb{1} \otimes \sigma)$ \\ 
     \hline
     Final System State & $\rho_1 = \frac{1}{2} [\mathbb{1} + c^2 \vec{\beta} \cdot \vec{\sigma} + s^2 \vec{\alpha} \cdot \vec{\sigma} $ \\
     & $ + \frac{cs}{4} (\vec{\beta} \times \vec{\alpha}) \cdot \vec{\sigma}]$ \\
     \hline
    \end{tabular}
    \label{state_table2}
    \end{table}
 
    Our incoherent quantum homogenization protocol involves a system qubit, a reservoir of identical environment qubits, and a set of control qubits. The system qubit interacts sequentially with each environment qubit through a \textsc{cswap} gate, moderated by a new control qubit in the state $\ket{c} = \cos \eta \ket{0} + \sin \eta \ket{1}$.

\begin{figure}[H]
        \includegraphics[width= \linewidth]{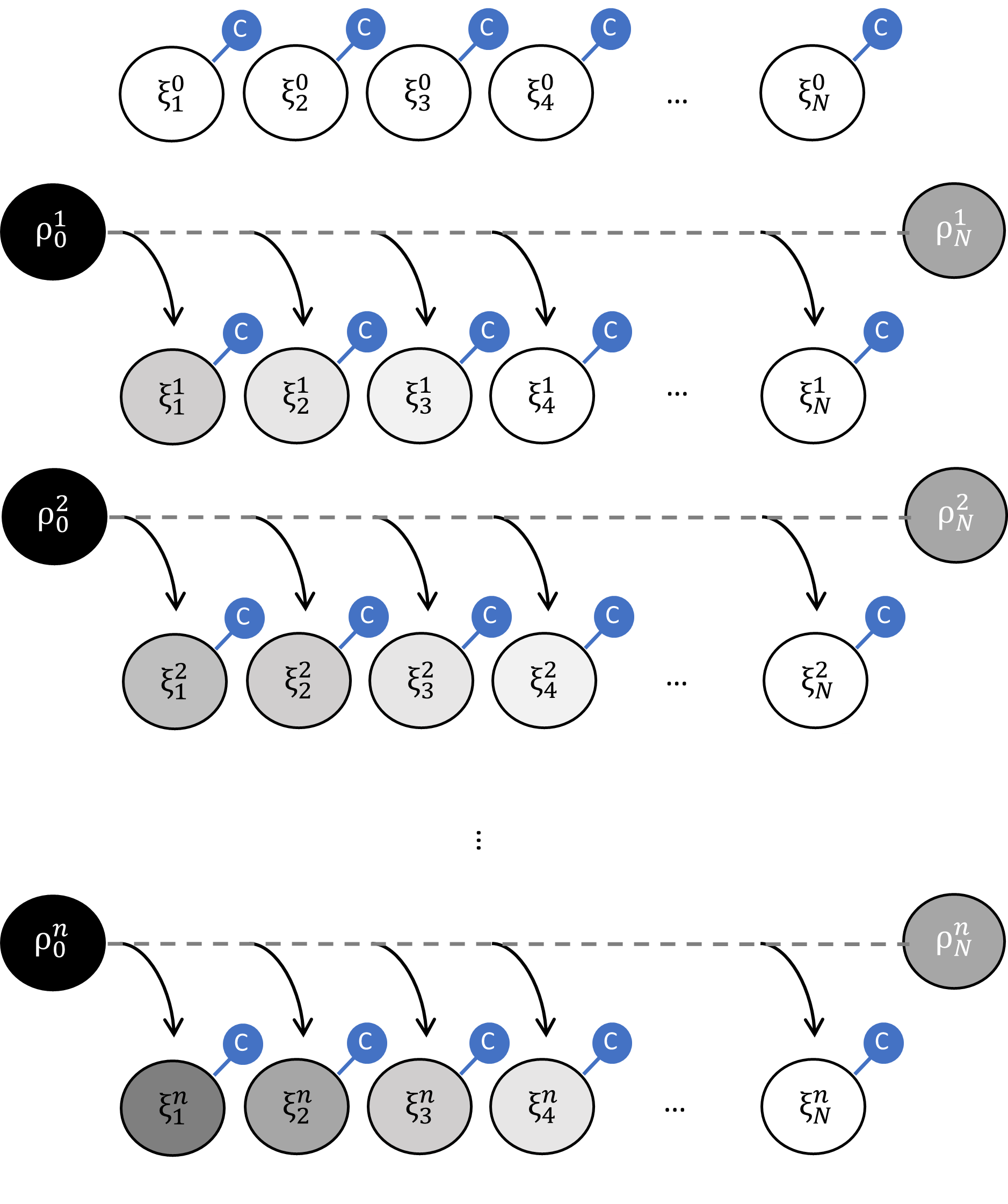}
        \caption{The \textsc{cswap} homogenizer. Control qubits are labelled c. The $j^{\textrm{th}}$ reservoir qubit after interaction with the $i^{\textrm{th}}$ system is in the state $\xi_j^i$, and the $i^{\textrm{th}}$ system after interaction with the $j^{\textrm{th}}$ reservoir qubit is in the state $\rho_j^i$.}
        \label{protocol_diagram}
    \end{figure}

    The protocol is shown in Figure \ref{protocol_diagram}, for a reservoir of $N$ qubits. The initial state of the reservoir qubits is the target final state for the system qubit. \textsc{cswap} operations between the system and reservoir qubit, moderated by a control qubit, are represented by an arrow. The homogenization of multiple systems is considered in Section \ref{reusability}. First we consider the homogenization of a single system.
    
\section{Convergence}\label{convergence}

    Here we demonstrate that the \textsc{cswap} homogenizer has the same convergence properties as the \textsc{pswap} homogenizer, meaning that convergence is not affected by the coherence terms. 
    
    \subsection{State Fidelity}\label{fidelity}
        
        We will show that the two homogenizers achieve the same convergence using fidelity as a measure of distance, first using analytic calculations and supported by a Qiskit simulation \cite{ibmq}. 
        
        The aim of a homogenization protocol is to approximate $F(\rho_N,\xi) = 1$ as closely as possible, where $\rho_N$ is the system qubit state after interacting with $N$ reservoir qubits, $\xi$ is the initial state of the reservoir qubits, and $F(\rho_N,\xi)$ is the fidelity between $\rho_N$ and $\xi$.
        
        For a system starting with Bloch vector $\vec{\beta}$ and reservoir qubit with Bloch vector $\vec{\alpha}$, with the shorthand $c = \cos \eta$, $s = \sin \eta$, Table \ref{state_table1} shows that for the \textsc{cswap}:
        \begin{equation}
            \vec{\beta}_1 = c^2\vec{\beta} + s^2\vec{\alpha},
        \end{equation}
        
        and for the \textsc{pswap}
        \begin{equation}
            \vec{\beta}_1 = c^2\vec{\beta} + s^2\vec{\alpha} + \frac{cs}{4} \vec{\beta} \times \vec{\alpha},
        \end{equation}

        where the subscript indicates that the system qubit has interacted with one reservoir qubit.
        
        The fidelity between the system and reservoir state $\vec{\alpha}$ for the incoherent homogenizer, using the \textsc{cswap}, is 
        \begin{equation} \label{cswap_fid}
            F_{inc} = \frac{1}{2}(1+c^2 \vec{\beta} \cdot \vec{\alpha} + s^2) + \frac{1}{2} \sqrt{(1-|c^2\vec{\beta} + s^2\vec{\alpha}|^2)(1 - |\vec{\alpha}|^2)}
        \end{equation}
        and for the coherent homogenizer, using the \textsc{pswap}, is
        \begin{align*}
            F_{coh} = \frac{1}{2}(1+c^2 \vec{\beta} \cdot \vec{\alpha} + s^2) + ~~~~~~~~~~~~~~~~~~~~~~~~~~~~~~~~~~~~~~~~~~
        \end{align*}
        \begin{equation}\label{pswap_fid}
            \frac{1}{2} \sqrt{(1-|c^2\vec{\beta} + s^2\vec{\alpha} + \frac{cs}{4} \vec{\beta} \times \vec{\alpha}~|^2)(1 - |\vec{\alpha}|^2)}
        \end{equation}

\begin{figure}[H]
            \includegraphics[width=\linewidth]{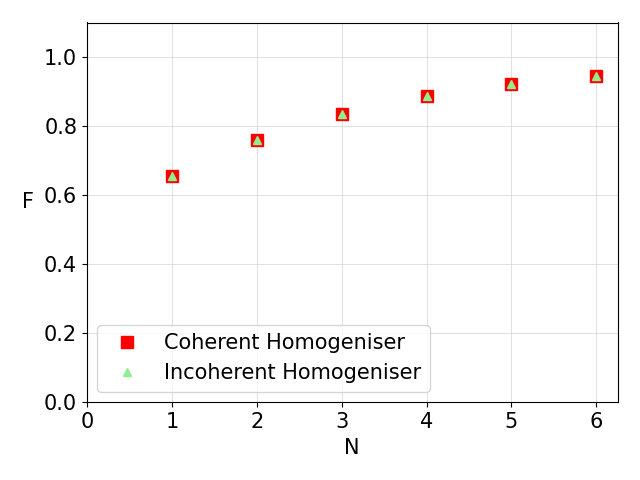}
            \caption{State fidelity against number of system-reservoir interactions for transforming $\ket{0}$ to $\ket{+}$.}
            \label{fidelitygraph}
        \end{figure}

        The additional term introduced in the \textsc{pswap} fidelity is zero if $|\vec{\alpha}| = 1$, $\vec{\beta} \parallel \vec{\alpha}$, $\vec{\alpha} = 0$ or $\vec{\beta} = 0$. Even at its maximum, the additional term has a significantly smaller contribution to the fidelity than the other terms. Specifically, in Appendix \ref{comparison} we derive an upper bound on the difference between the fidelities: 

\begin{equation}
\frac{\delta F}{F_{inc}} \leq \frac{\sqrt{1-\alpha^2} (\sqrt{3 - \alpha^2} - \sqrt{3 - \alpha^2 - \frac{\alpha}{2}})}{3 + \sqrt{1-\alpha^2}\sqrt{3 - \alpha^2}}
\end{equation}

where $\delta F = F_{inc} - F_{coh}$, with the maximum difference being by a factor of approximately $2\%$. Furthermore, the difference in fidelity tends towards zero when additional \textsc{cswap} and \textsc{pswap} gates are applied and the size of the reservoir used for homogenization is increased. This is due to the additional term in the \textsc{pswap} fidelity being scaled by a factor that tends towards zero as the Bloch vectors of the system and reservoir qubits converge to become parallel with more interactions with the reservoir, also discussed in Appendix \ref{comparison}. Hence, the convergence properties of the fidelities, for a large reservoir size, will be equivalent. Therefore there is a close agreement between the state fidelity outcomes achieved by the two homogenization protocols, for arbitrary system and reservoir states. A Qiskit simulation of both protocols transforming a system originally in the $\ket{0}$ state to the $\ket{+}$ state is shown in Figure \ref{fidelitygraph}. 

        The incoherent \textsc{cswap} homogenizer therefore achieves the same accuracy as the coherent \textsc{pswap} homogenizer, up to a small correction which tends to zero in the limit of a large homogenizer. This demonstrates that the coherence introduced by the \textsc{pswap} is not contributing to the homogenization properties.

    \subsection{Trace Distance}\label{tracedistance}
    Another way of demonstrating an equivalence between the homogenizers is calculating the minimum number of system-reservoir interactions required so that the trace distance between the final system state and original reservoir state is below some error:
    
    \begin{equation}\label{s_delta}
        D(\rho_{N},\xi) \leq \delta 
    \end{equation}
    
    whilst also having the distance of every environment state with the original reservoir state being below that error: 
    
    \begin{equation}\label{e_delta}
        D(\xi_i, \xi) \leq \delta ~ \forall i, i \leq N.
    \end{equation}

    Here $\rho_{N}$ is a system state that has interacted with $N$ reservoir qubits, $\xi$ is the original reservoir qubit state, and $\xi_i$ is the state of the $i^{\textrm{th}}$ reservoir qubit. Following the method in \cite{2002paper} we use the trace distance as a measure of the distance between two states, corresponding to the distance on the Bloch sphere between the two states' Bloch vectors.
    
    Since $\xi_1$ interacts first with the system qubit it will be furthest from the reservoir state, so as long as this satisfies Equation \ref{e_delta} all other reservoir qubits also satisfy Equation \ref{e_delta}. Using Table \ref{state_table1}:
        \begin{equation}
            \xi_1 = \frac{\mathbb{1}}{2} +  \frac{s^2}{2} \vec{\beta} \cdot \vec{\sigma} + \frac{c^2}{2} \vec{\alpha} \cdot \vec{\sigma}.
        \end{equation}

    We will consider the initial system state and initial reservoir states having an absolute difference between their Bloch vectors given by: 
        
        \begin{equation}
            d = |\vec{\beta} - \vec{\alpha}|.
        \end{equation}

    This makes our analysis initially more general than the bound derived in \cite{2002paper}, where the extreme case of a distance between the states of 2 (where the initial system and reservoir states are orthogonal) is assumed from the beginning. The relevant trace distance for the \textsc{cswap} case is simpler analytically than the \textsc{pswap}: for \textsc{cswap}, the cross-terms in Table \ref{state_table1} are zero for any $d$, but for the \textsc{pswap}, the cross-terms in Table \ref{state_table2} are zero for $d=2$ (initially orthogonal qubits) but not in general. Hence, we can use the general distance $d$ to find tighter bounds on the resources needed for the \textsc{cswap} protocol. Later we will specialise to $d=2$ to compare to the \textsc{pswap} homogenizer results for the worst-case homogenization. In our general \textsc{cswap} case, the trace distance between the first reservoir qubit after it has interacted with the system with its original state is:  

    \color{black}
        
        \begin{equation}
            D( \xi_1, \xi ) = d s^2.
        \end{equation}
        So the limit for satisfying Equation \ref{e_delta} is
        \begin{equation}
             s^2 = \frac{\delta}{d}.
        \end{equation} 
        
        The system state after N system-reservoir interactions is
        \begin{equation}
            \rho_N = \frac{\mathbb{1}}{2} + \frac{c^{2N}}{2}  \vec{\beta} \cdot \vec{\sigma} + \frac{(1 - c^{2N})}{2}  \vec{\alpha} \cdot \vec{\sigma},
        \end{equation}
        so that
        \begin{equation}\label{1}
            D( \rho_N, \xi ) = d c^{2N}.
        \end{equation}
        Using $s^2 = \frac{\delta}{d}$ we get
        \begin{equation}
            D( \rho_N, \xi ) = d \left[1 - \frac{\delta}{d}\right]^N.
        \end{equation}
        To satisfy Equation \ref{s_delta} we require 
        \begin{equation}
            D( \rho_N, \xi ) \leq \delta.
        \end{equation}
        Solving for N, 
        \begin{equation} \label{NBound}
            N \geq \frac{\ln \frac{\delta}{d}}{\ln(1 - \frac{\delta}{d})}.
        \end{equation}
        This is the minimum number of gates required to achieve convergence to within $\delta$ for a system qubit and all the reservoir qubits.
        
        In \cite{2002paper} it is shown that for the \textsc{pswap} homogenizer the number of gates required to achieve convergence within $\delta$ for the case of two orthogonal pure states is
        
        \begin{equation}
            N_{\delta} \geq \frac{\ln \frac{\delta}{2}}{\ln(1 - \frac{\delta}{2})}.
        \end{equation}
        
        This is the same as our result for the \textsc{cswap} homogenizer, where for orthogonal pure states 
        
        \begin{equation}
        d = |\vec{\beta} - \vec{\alpha}| = 2,
        \end{equation}
        so that 
        \begin{equation}
                N \geq \frac{\ln \frac{\delta}{2}}{\ln(1 - \frac{\delta}{2})}.
            \end{equation}
        
        Therefore we have derived an equivalent upper bound on the number of reservoir qubits needed for a successful homogenization using the \textsc{cswap} homogenizer as for the \textsc{pswap} homogenizer.

    \subsection{Differences between homogenizers}

    Despite the similarity in convergence properties of the two homogenizers, we also found significant differences between them in how the Bloch vector of the system qubit evolves during homogenization, and in the joint system-reservoir von Neumann entropy. We explain these differences and demonstrate them using simulations in Appendix \ref{appendix_differences}.
        
    \section{Repeated homogenization}\label{reusability}
            
            Now we will consider the reusability of both homogenizers, which was not considered in \cite{2002paper}. Reusability is of interest in quantum thermodynamics for assessing whether transformations can be enabled using catalysts. For instance, analysis using resource theories has found that catalysts can drastically increase performable state transformations \cite{ng2015limits, ng2019resource, brandao2015second, lipka2021all}. 
            Furthermore, reusability is important for assessing whether a transformation can be performed reliably, as in the Constructor Theory approach to thermodynamics \cite{deutsch2013constructor, marletto2016constructor, marletto2022information}. In Appendix \ref{appendix_erasure}, we discuss how the similarity we derived earlier between the \textsc{pswap} and \textsc{cswap} homogenizers shows how those results can be generalized to a wider class of incoherent protocols.

            Here we use a different approach to investigating the reusability of homogenization machines, bounding the number of reservoir qubits $N$ required and number of times $n$ the homogenization task can be performed, in order to satisfy the conditions for homogenization. For a reusable homogenizer, we require that all homogenizer qubits remain arbitrarily close to their original states, but we extend the requirement on the system qubit such that all $n$ system qubits must be homogenized arbitrarily close to the original reservoir qubits' state. 
            
            For both the \textsc{cswap} and \textsc{pswap}, we find that there is a finite number of homogenizer qubits required such that a given homogenizer is able to transform a given number of system qubits from an initial state to a final state, within a specified error, with all homogenizer qubits also being within a certain error from their original state. This is a natural extension of the conditions for homogenization introduced in \cite{2002paper} to a setting where the homogenizer needs to also be reused to transform some finite number of systems. 

            Note that for the \textsc{cswap} homogenizer, after the first system qubit has been homogenized the control qubits are reordered, so that for the next system qubits no reservoir qubit interacts with the same control. We show in Appendix \ref{appendix_derivations} that this prevents quantum interference terms developing between control and reservoir qubits. 
            
            The first reservoir qubit deteriorates most rapidly as this is first to interact with each new system qubit, so when this state is within the required distance from $\xi$ so are the other reservoir qubits.
            
            The state of the first reservoir qubit after $n$ interactions with a fresh system qubit is
            \begin{equation}
                \vec{\alpha}_1^n = (1 - c^{2n}) ~ \vec{\beta} + c^{2n} \vec{\alpha_1^0}
            \end{equation}

            where the superscript indicates the number of system qubits the reservoir qubit has interacted with.
        
            As in the previous section, for homogenization we require $D( \xi^{n}_{1}, \xi ) \leq \delta$. Hence, $1 - c^{2n} \leq \frac{\delta}{d}$, such that
            
            \begin{equation}\label{reuse_constraint}
                c^{2n} \geq 1 - \frac{\delta}{d}.
            \end{equation}
           This can be rearranged to find the constraint on the number of qubits that can be homogenized for a given error and coupling strength: 

            \begin{equation}
                n \leq \frac{\textrm{ln}(1-\frac{\delta}{d})}{2 \textrm{ln(c)}}.
            \end{equation}

            Now we can consider constraining the system qubits such that every system qubit is within a distance $\epsilon$ from the most-deteriorated reservoir qubit, namely the first one, which is in state $\xi^{(n-1)}_1$ before the final homogenization. Therefore in a worst-case scenario, we could use a homogenizer entirely composed of qubits in the state of the most deteriorated one, $\xi_1^{(n-1) \otimes N}$, to transform the state of a system qubit. The lower bound on the number of reservoir qubits needed for the system qubit to be within a distance $\epsilon$ of the reservoir qubits' states is of the same form as the original bound when the homogenizer was used once, in Equation \ref{NBound}:

            \begin{equation} \label{NBound2}
                N \geq \frac{\ln \frac{\epsilon}{d'}}{\ln(1 - \frac{\epsilon}{d'})}.
            \end{equation}

            Here $d' = |\vec{\beta}^0 - \vec{\alpha}^{(n-1)}_1|$ is the distance between the first reservoir qubit after interacting with $n-1$ system qubits, and the original system qubit state on the Bloch sphere. Using $d = |\vec{\beta}^0 - \vec{\alpha}^0_1|$, we find $d - d' = |\vec{\alpha}^0_1 - \vec{\alpha}^n_1|$. Simplifying this expression leads to the following relation between $d$ and $d'$: 

            \begin{equation}
                d' = c^{2n}d.
            \end{equation}
            
            Now the distance of the worst-case reservoir qubit $\xi^{(n-1)}_1$ from the target state is $d(1-c^{2(n-1)})$, from Equation \ref{reuse_constraint}. Therefore the distance $\epsilon$ must satisfy the condition $\Delta = \epsilon + d(1-c^{2(n-1)})$, for all system qubits to be within $\Delta$ of the target state. Substituting the resulting expression for $\epsilon$ into Equation \ref{NBound2}, along with the expression for $d'$, the bound can be rewritten as: 

            \begin{equation} \label{NBound3}
                N \geq N_{\min} = \frac{\ln(1 - \frac{d-\Delta}{dc^{2(n-1)}})}{\ln \frac{d-\Delta}{dc^{2(n-1)}}}.
            \end{equation}

            Now if the conditions in Equation \ref{NBound3} and Equation \ref{reuse_constraint} are both satisfied, then $N$ reservoir qubits and $n$ system qubits are a maximum distance $\Delta$ from the original reservoir qubits' state. Since the bound comes from a worst-case approximation, the minimum $N$ needed for specific transformations will be smaller than $N_{\min}$. We can therefore always homogenize $n$ qubits, with all system and reservoir qubits within an error $\Delta$, for any $n$ and $\Delta$, by making $\eta$ sufficiently small and $N$ sufficiently large. For the single-use homogenizer, reducing the desired error $\Delta$ requires $\eta$ to decrease and $N$ to increase. For our reusable homogenizer, we have the added condition that imposing $n$ to be greater also requires $\eta$ to decrease and $N$ to increase, further constraining the conditions for homogenization. Note that setting $n=1$ and $d=2$ reproduces the constraints on $\eta$ and $N$ derived in \cite{2002paper}. 

            We derived the conditions on $N$ and $n$ for a general initial distance $d$ between the initial system qubit state and initial homogenizer qubits' state. This general expression holds for the \textsc{cswap} homogenizer. By considering the worst-case scenario where the initial reservoir state and initial system states are a distance $d=2$ apart (orthogonal pure states), then we have conditions for reusable partial swap homogenization.

\section{Conclusion}

    We proposed a model for a universal quantum homogenizer that does not have coherence between the system and reservoir qubits, based on a \textsc{cswap} operation instead of \textsc{pswap}. We computed an upper-bound on the difference between the reduced states of the system and reservoir qubits of the \textsc{cswap} homogenizer compared to the \textsc{pswap}, showing that it tends to zero in the limit of a large reservoir, and simulated an example where the homogenization protocols are equivalent. Then we derived a bound on the resources needed for an arbitrarily good \textsc{cswap} homogenization, showing that it satisfies the required convergence conditions for homogenization. Our result is more general than that previously derived for the \textsc{pswap}, showing the dependence of resources required on the distance between the initial system and reservoir qubit states. We also contrasted the \textsc{cswap} and \textsc{pswap} homogenizers in terms of the von Neumann entropy of the joint system-reservoir qubits. 

    Then we analysed how far the coherent and incoherent homogenizers can be re-used to perform state transformations, deriving constraints on the resources needed to repeatedly perform imperfect homogenizations. Our analysis also suggests that recent demonstrations of a new kind of irreversibility based on homogenization machines can be generalised to incoherent models for thermalization and information erasure. 

    Future work could investigate connections between the general bounds on repeated homogenizations found here with approaches to modelling catalysts in quantum resource theories. Another interesting avenue is to investigate in more detail how entanglement builds up in the two homogenizers, building on recently-proposed approaches to describe quantum correlations in collision models (e.g. \cite{sergey2022entanglement}). Entanglement may be distributed differently in the coherent and incoherent homogenizers, despite the negligible differences in the ultimate convergence properties. 

\section*{Acknowledgements}

MV is grateful to the Heilbronn Institute for Mathematical Research for their support. This research was made possible through the generous support of the Gordon and Betty Moore Foundation and the Eutopia Foundation. 
    
\printbibliography

\appendix
     
\section{Controlled Swap Derivations}\label{appendix_derivations}

        Here we derive the reduced states of the system qubit and reservoir qubit after interacting via a \textsc{cswap} operation.

        Let the starting states of a control qubit, system qubit and reservoir qubit be $\rho_c^i$, $\rho_s^i$ and $\rho_r^i$ respectively, with Bloch vectors $\vec{c}$, $\vec{s}$ and $\vec{r}$. We have initial states:
        
        \begin{equation}
            \rho_c^i = \frac{\mathbb{1} + \vec{c} \cdot \vec{\sigma}}{2},
        \end{equation}
        
        \begin{equation}
            \rho_s^i = \frac{\mathbb{1} + \vec{s} \cdot \vec{\sigma}}{2},
        \end{equation}
        
        \begin{equation}
            \rho_r^i = \frac{\mathbb{1} + \vec{r} \cdot \vec{\sigma}}{2}.
        \end{equation}
        
        Then let the controlled swap operator be $U$ from Equation \ref{control swap}, and act on these states: 
        \begin{equation}
            U \{ \rho_c \otimes \rho_s \otimes \rho_r \} U ^\dag .
        \end{equation}
        
        We then obtain final states of $\rho_c^f$, $\rho_s^f$ and $\rho_r^f$, where $c_x$, $c_y$ and $c_z$ are the x, y and z components of the Bloch vector $\vec{c}$ :
        
        \begin{equation}\label{3}
            \rho_c^f = \frac{\mathbb{1}}{2}  +  \left( 1 + \vec{r} \cdot \vec{s} \right) \frac{c_x \sigma_x}{4} + \left( 1 + \vec{r} \cdot \vec{s} \right) \frac{c_y \sigma_y}{4} + \frac{c_z \sigma_z}{2},
        \end{equation}
        
        \begin{equation}
            \rho_s^f = \frac{\mathbb{1}}{2} + \frac{(1-c_z)}{4} \vec{e} \cdot \vec{\sigma} + \frac{(1+c_z)}{4} \vec{s} \cdot \vec{\sigma},
        \end{equation}
        
        \begin{equation}
            \rho_r^f = \frac{\mathbb{1}}{2} + \frac{(1+c_z)}{4} \vec{r} \cdot \vec{\sigma} + \frac{(1-c_z)}{4} \vec{s} \cdot \vec{\sigma}.
        \end{equation}
        
        Since the final states of system and reservoir depend only on 
        \begin{math}
        c_z
        \end{math}
        we can let 
        \begin{math}
        c_z = 2\cos^2{\eta} - 1
        \end{math}
        so that the controlled swap is parameterized by the swap strength 
        \begin{math}
        \eta
        \end{math}
        \cite{2002paper}. 
        
        Then we see the system and reservoir final states can be written simply as 
        \begin{equation}
            \rho_s^f = \frac{\mathbb{1}}{2} + \frac{\sin^2{\eta}}{2} \vec{r} \cdot \vec{\sigma} + \frac{\cos^2{\eta}}{2} \vec{s} \cdot \vec{\sigma},
        \end{equation}
        
        \begin{equation}
            \rho_r^f = \frac{\mathbb{1}}{2} + \frac{\cos^2{\eta}}{2} \vec{r} \cdot \vec{\sigma} + \frac{\sin^2{\eta}}{2} \vec{s} \cdot \vec{\sigma}.
        \end{equation}
    
        Also, because the final states of the system and reservoir only depend on $c_z$, which remains unchanged after a \textsc{cswap} as shown in Equation \ref{3}, the control can be reused as long as it is with a different reservoir and system qubit each time, avoiding interference terms between the control and its target qubits. Note that this relies on the number of reservoir qubits being greater than the number of system qubits being homogenized for there to be different control qubits used in each interaction, which is consistent with a large reservoir being the typical regime in which homogenization is used to transform system states. 

        The final states after one \textsc{cswap} interaction are summarized in Tables \ref{state_table1} and \ref{state_table2}. 

\section{Bounding fidelity difference} \label{comparison}

Here we bound the difference between the magnitudes of the system qubit's fidelity with the target state in the \textsc{pswap} and \textsc{cswap} homogenizers, hence the accuracy of the homogenization. We show that the ratio of the magnitudes of the additional term in the \textsc{pswap} fidelity to the \textsc{cswap} fidelity is much less than one. The ratio is: 

\begin{equation}
\frac{\delta F}{F_{inc}} = \frac{F_{inc} - F_{coh}}{F_{inc}},
\end{equation}

where        

\begin{equation}
            F_{inc} = \frac{1}{2}(1+c^2 \vec{\beta} \cdot \vec{\alpha} + s^2) + \frac{1}{2} \sqrt{(1-|c^2\vec{\beta} + s^2\vec{\alpha}|^2)(1 - |\vec{\alpha}|^2)}
        \end{equation}
        and 
        \begin{align*}
            F_{coh} = \frac{1}{2}(1+c^2 \vec{\beta} \cdot \vec{\alpha} + s^2) + ~~~~~~~~~~~~~~~~~~~~~~~~~~~~~~~~~~~~~~~~~~
        \end{align*}
        \begin{equation}
            \frac{1}{2} \sqrt{(1-|c^2\vec{\beta} + s^2\vec{\alpha} + \frac{cs}{4} \vec{\beta} \times \vec{\alpha}~|^2)(1 - |\vec{\alpha}|^2)}
        \end{equation}, repeating Equations \ref{cswap_fid} and \ref{pswap_fid} here for clarity. 

To bound the maximum value of this ratio, we consider the case where $\vec{\alpha}$ is perpendicular to $\vec{\beta}$, such that $\vec{\alpha} = \alpha \hat{z}$, $\vec{\beta} = \beta \hat{x}$, and $\vec{\beta} \times \vec{\alpha} = \alpha \beta \hat{y}$, maximising the difference between the two fidelities. 

Then the upper bound on $\frac{\delta F}{F_{inc}}$ is: 

\begin{equation}
\frac{\sqrt{1-\alpha^2} \left( \sqrt{1-c^4 \beta^2 + s^4 \alpha^2} - \sqrt{1-c^4 \beta^2 + s^4 \alpha^2 - \frac{c^2 s^2 \alpha \beta}{2}}\right)}{1 + s^2 + \sqrt{1-\alpha^2}\sqrt{1-c^4 \beta^2 + s^4 \alpha^2}}.
\end{equation}

From the form of the extra term in the coherent fidelity, the difference will be maximised for $\beta=1$ and $c = s = \frac{1}{\sqrt{2}}$. Then we can further simplify the bound, solely in terms of $\alpha$:

\begin{equation}
\frac{\delta F}{F_{inc}} \leq \frac{\sqrt{1-\alpha^2} (\sqrt{3 - \alpha^2} - \sqrt{3 - \alpha^2 - \frac{\alpha}{2}})}{3 + \sqrt{1-\alpha^2}\sqrt{3 - \alpha^2}}.
\end{equation}

The maximum of the RHS as $\alpha$ varies between $0$ and $1$ is $\approx 0.0208$ at $\alpha \approx 0.805$. Hence, at a maximum, $F_{coh}$ has an approximately $2\%$ deviation from $F_{inc}$.

Now let's consider the effect that subsequent interactions of the homogenization protocols have on this difference in fidelity. For our worst-case upper-bound we initialised $\vec{\alpha}$ and $\vec{\beta}$ to be perpendicular. From the convergence properties of the $\textsc{pswap}$, for subsequent interactions with reservoir qubits, the Bloch vectors of the system and reservoir states will no longer be perpendicular and will tend towards the same direction. The deviation between the coherent and incoherent fidelities will be scaled down by a factor of $\textrm{sin}{\theta}$ due to the contribution from the cross-product of the vectors $\vec{\alpha}$ and $\vec{\beta}$, where $\theta$ is the angle between the system and reservoir qubit Bloch vectors. This scaling factor will tend towards zero, the more reservoir interactions are included in the homogenization protocol. This means that the convergence properties of the $\textsc{pswap}$ and $\textsc{cswap}$ fidelities in this limit are equivalent. 

In summary, there is a $\approx 2\%$ upper bound on the fidelity deviation of the coherent homogenizer from the incoherent homogenizer for a finite number of system interactions with the reservoir, and in the limit of a large reservoir, the difference between the fidelities tends towards zero. 

\section{Differences between homogenizers}\label{appendix_differences}

\subsection{Evolution of Bloch vectors}\label{appendix_bloch}

\begin{figure}[H]
            \includegraphics[width=\linewidth]{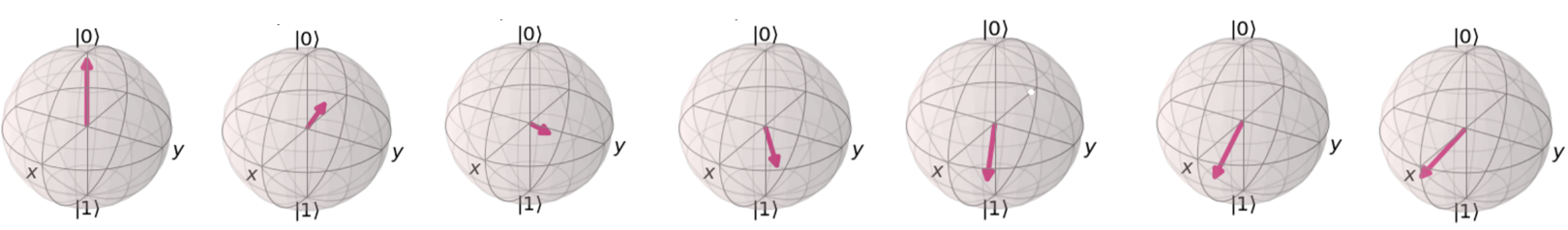}
            \caption{System Bloch vector evolution for the coherent homogenizer with initial state $\ket{0}$ and reservoir state $\ket{+}$, simulated using Qiskit \cite{ibmq}.}
            \label{pswap_series}
        \end{figure}

\begin{figure}[H]
            \includegraphics[width=\linewidth]{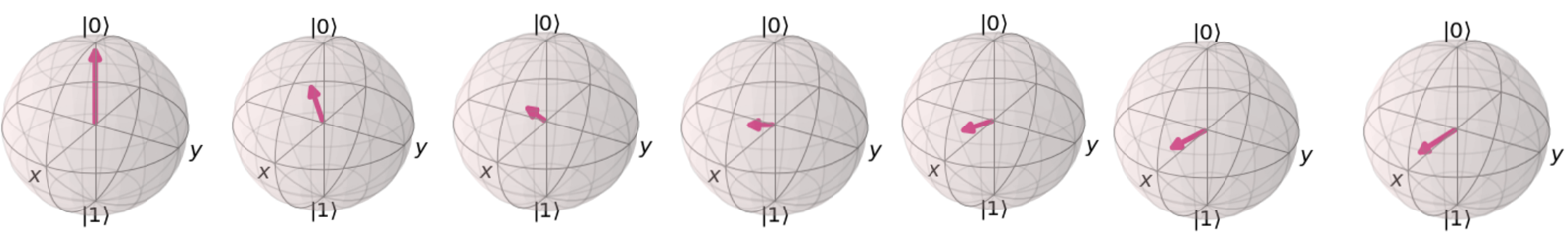}
            \caption{System Bloch vector evolution for the incoherent homogenizer with initial state $\ket{0}$ and reservoir state $\ket{+}$, simulated using Qiskit \cite{ibmq}.}
            \label{cswap_series}
        \end{figure}

        Despite the similarity in fidelities computed for the two homogenization protocols, Figures \ref{pswap_series} and \ref{cswap_series} show that there is a significant difference in how the states are evolving on the Bloch sphere. In the \textsc{cswap} case, the Bloch vector remains in the X-Z plane throughout its evolution. The coherence term in the \textsc{pswap} case changes the path the Bloch vector takes but not the final state.


\subsection{Joint system-reservoir entropy}\label{appendix_entropy}

    Here we show the significant difference in joint von Neumann entropy of the system and reservoir qubits for the \textsc{pswap} and \textsc{cswap} homogenizers, which nonetheless does not affect the homogenization properties. Specifically the joint von Neumann entropy is $S = - tr(\rho_{s+r} \log \rho_{s+r})$ where $\rho_{s+r}$ is the joint state of the system and reservoir qubits (with the control qubit traced out for the \textsc{cswap}).
        
    With the coherent \textsc{pswap} homogenizer, all interactions between the system and reservoir qubits are unitary, and hence the overall von Neumann entropy is constant. By contrast, the incoherent \textsc{cswap} homogenizer involves a control qubit which is traced out to find the joint system and reservoir state. Therefore we expect that the system-reservoir von Neumann entropy in general changes with number of interactions. Specifically, since the entanglement of the system-reservoir qubits with the control qubit contributes negatively to the von Neumann entropy, we might intuitively expect that the joint system-reservoir von Neumann entropy increases with number of interactions. 

\begin{figure}[H]
            \includegraphics[width=\linewidth]{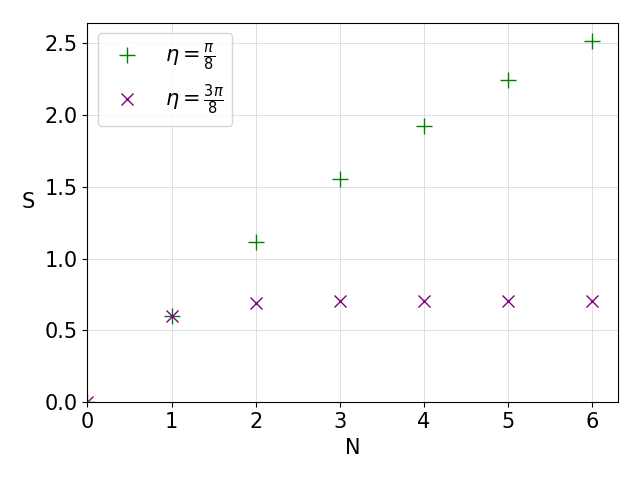}
            \caption{Von Neumann entropy as a function of the number of system-environment interactions with coupling strengths $\eta = \frac{\pi}{8}$ and $\eta = \frac{3\pi}{8}$.}
            \label{entropy}
        \end{figure}

    When we compute numerical simulations of the von Neumann entropy for the joint system-reservoir state with \textsc{cswap} interactions, we indeed find that it increases with interactions, and then reaches a plateau, which happens sooner for strong coupling than weak coupling, though at a smaller value of maximum von Neumann entropy. This can be understood in terms of the system being homogenized quicker in the strong coupling case (leading to a plateau in joint system-reservoir von Neumann entropy) but there is also more negative entropy contributed by the entanglement with the control qubit (leading to a smaller maximum value of von Neumann entropy), shown in Figure \ref{entropy}.  
        
\section{Comparison to other reusability results}\label{appendix_erasure}

    A pertinent question is whether the convergence and irreversibility properties of the quantum homogenizer found in \cite{marletto2022emergence, violaris2022irreversibility} are dependent on the web of interference between system and reservoir qubits that arises due to the coherent partial swap interactions. Those works demonstrate an asymmetry in reusing homogenization machines to transform qubits from mixed to pure states and the opposite process. If the coherence is important, this would suggest that the homogenizer's properties are a non-trivial quantum effect, whereas if the properties are independent of the coherence, this suggests the homogenizer's properties can be generalized to a wider class of incoherent protocols, which are closer to classical implementations of thermalization and information erasure. The similarity we derived between the \textsc{pswap} and \textsc{cswap} homogenizers supports the latter case.

    The results in \cite{marletto2022emergence, violaris2022irreversibility} were derived using a quantity called the relative deterioration, which is a function of fidelities between qubits and their target states. We showed in Section \ref{fidelity} that the difference between the \textsc{pswap} and \textsc{cswap} fidelities of system states with the target state is small and tends to zero in the limit of a large reservoir. Since relative deterioration is quantified using fidelities, the equivalence of the convergence properties of the homogenizers suggests that the additional repeatability cost of coherently erasing information also applies for the incoherent model. Hence our work expands the range of applicability of the repeatability cost of erasure to a wider class of models. By contrast with protocols in recent studies of coherence as a resource for quantum information processing tasks \cite{streltsov2017colloquium}, it is not a resource for the homogenization and information erasure protocols presented here. 

    We also note that the bounds we derived on $\eta$ and $N$ in this paper assumed a worst-case homogenization. Hence, we cannot directly use such bounds to compare the resource costs of more specific tasks, such as transforming pure states to mixed states and the opposite process. In addition, the asymmetry shown between pure and mixed state homogenization is in a different physical context to \cite{violaris2022irreversibility}, where the coupling strength $\eta$ is first fixed, and then it is considered how far the homogenization machines can be reused to perform a homogenization within a given error, while remaining close to their original state. 

\end{multicols}
\end{document}